\documentclass{aa}
\usepackage{amsmath}
\usepackage{natbib}
\usepackage{graphicx}
\usepackage{multicol}
\bibpunct{(}{)}{;}{a}{}{,} % to follow the A&A style

\begin{document}

     \title{Pulse energy distribution for RRAT J0139+33 according to observations at the frequency 111 MHz}
   %\titlerunning{Pulse energy distribution for RRAT J0139+33}
      \author{
E.A.Brylyakova
                    \and
S.A. Tyul'bashev}
   \institute{
Lebedev Physical Institute, Astro Space Center, Pushchino Radio Astronomy Observatory,\\
142290, Moscow region, Russia\\
  \email{serg@prao.ru}
          }
   \date{Received ; accepted }
  \abstract
         {Using five year monitoring observations, we did a blind search for pulses for rotating radio transient (RRAT) J0139+33 and PSR B0320+39. At the interval $\pm1.5^m$ of the time corresponding to the source passing through the meridian, we detected 39377 individual pulses for the pulsar B0320+39 and 1013 pulses for RRAT J0139+33. The share of registered pulses from the total number of observed periods for the pulsar B0320+39 is $74\%$, and for the transient J0139+33 it is $0.42\%$. Signal-to-noise ratio ($S/N$) for the strongest registered pulses is approximately equal to: $S/N=262$ (for B0320+39) and $S/N=154$ (for J0139+33).

               Distributions of the number of detected pulses in $S/N$ units for the pulsar and for the rotating transient are obtained. The distributions could be approximated with a lognormal and power dependencies. For B0320+39 pulsar, the dependence is lognormal, it turns into a power dependence at high values of $S/N$, and for RRAT J0139+33, the distribution of pulses by energy is described by a broken (bimodal) power dependence with an exponent of about 0.4 and 1.8 ($S/N\le 19$ and $S/N \ge 19$).

            We have not detected regular (pulsar) emission of J0139+33. Analysis of the obtained data suggests that RRAT J0139+33 is a pulsar with giant pulses.}


      \maketitle
%
%-------------------------------------------------------------------

\section{Introduction}

   The discovery of pulsars in 1967 was associated with the detection of regular (periodic) pulses of emission in the meter range of wavelengths. These pulses were similar with ``weak sporadic interference''
%   in appearance to the scintillations of a compact radio source on an interplanetary plasma
\citep{Hewish1968}. In 2006, pulsars were discovered that emitted irregular pulses \citep{McLaughlin2006}. These pulsars were called {\bf R}otating {\bf RA}dio {\bf T}ransients (RRAT). Time intervals between RRAT pulses can be from tens of seconds to hours. It is still not completely clear whether these pulsars are the same objects as canonical pulsars. The search for an evolutionary relationship between canonical pulsars and RRAT was considered in papers \citet{Keane2008, Keane2011}.

   %Two of them seem to us to be the most substantiated \citep{Weltevrede2006,Zhang2007}.

      There are several hypotheses with attempts to explain the irregularity of RRAT emission. According to \citet{Weltevrede2006}, RRAT can be ordinary pulsars with ``extreme bursts of radio emission''. If the distance to a pulsar is large, the sensitivity of the radio telescope is not sufficient to detect periodic emission and to obtain an average profile. At the same time, the energy of individual giant pulses may be high enough to detect them. According to another hypothesis \citet{Zhang2007}, RRAT are pulsars with long (extrime) nullings.
%      The phenomenon of nulling has long been known and consists of the fact that pulses from pulsars can be missing.  Different pulsars have different degrees of missed pulses. This degree is described by the fraction that represents the number of missed pulses to the total number of observed periods. RRAT can be pulsars with extreme nulling, i.e. pulsars for which the bulk of the pulses is missing.
   In the work \citet{Burke-Spolaor2010}, it is assumed that the duration of the "fading" window of pulsars increases in proportion to their lifetime. Thus, according to this hypothesis, RRAT  represents a late evolutionary phase in the pulsar's life. This work also shows that the hypothesis of giant pulses is not applicable for at least some of the observed fast radio transients. For example, the source J0941-39 is during part of the time observed as RRAT, and part of the time as pulsar. In the paper \citet{Wang2007}, it is assumed that RRAT are an extreme types of mode-switching pulsars. In the papers \citet{Li2006,Luo2007}, it is assumed that RRATs are extinct pulsars that are turned of the interaction of the pulsar's magnetosphere with the incident matter or the interaction of the magnetosphere with the surrounding matter.

      Thus, the only certain thing is that RRAT are a subclass of pulsars that are detected by the emission of their individual pulses, but the reason for this sporadic emission remains unclear. It is possible that different hypotheses are valid for different RRAT, and that RRAT is an intermediate class to different types of known pulsars. There are few full-fledged RRAT studies, and therefore it is not possible to discard any hypotheses yet.

      RRAT research is difficult. Because the appearance of pulses is unpredictable, and the time interval between detected pulses can reach up to several hours, a long-time observation is necessary for their statistic accumulation. It is known that to obtain an average pulsar profile, which is one of the basic characteristics of a pulsar, you need to accumulate several hundred, or better, several thousand pulses \citep{Lorimer2004}. If the RRAT emits an average of one pulse per minute, then it takes more than 16 hours of continuous observations to accumulate 1000 pulses and obtain an average profile. Until now, RRAT detection takes place on the world's best telescopes, and allocation of dozens of hours for a full-fledged study of even one RRAT is an expensive task.

      Since August 2014 on the LPA LPI radio telescope (Large Phased Array of P.N.Lebedev Physical Institute of the Russian Academy of Sciences), located in Pushchino, Moscow region,  round-the-clock observations are carried out, including those used for searching for pulsars \citep{Tyulbashev2016}. Processing of semi-annual data revealed 33 new RRAT's \citep{Tyulbashev2018a, Tyulbashev2018b}. For one of these rotating transients (RRAT J0139+33, other name PSR J0139+3336 \citep{Michilli2020}) a lot of pulses were detected in the blind search. J0139+33 has high peak flux densities in individual pulses. Therefore, we choose this transient to test the hypotheses about the nature of RRAT.

\section{Observations and data processing}
	
LPA LPI  is a Meridian-type radio telescope with a filled aperture. It is a phased array. The telescope's receiving elements consist of 16384 half-wave dipoles. The size of the antenna is 187 m in the East-West direction and 384 m  in the North-South direction. After a major upgrade of the antenna, its central frequency is 110.3 MHz, and its full band is  2.5 MHz. Several independent radio telescopes were created on the basis of a single antenna field. Details about the antenna upgrade and the main observation programs for its multibeam implementation are given in the works \citet{Shishov2016,Tyulbashev2016}.
 	
Observations of the monitoring program are carried out simultaneously in 96 beams of the LPA LPI antenna. The data are recorded by a 32-channel digital recorder with a 78 kHz channel bandwidth and readout time $\tau = 12.5$~ms. The size of the beam is approximately $0.5\degr \times 1\degr$. The instant observation area of LPA when operating in the monitoring mode is about 50 square degrees. An area of 17000 square degrees is viewed per 24 hours, and each point in the sky is observed once in 24 hours for 3.5 minutes at half the power of the beam when passing through the meridian at a declination of zero degrees. The accumulated total observation time at each point in the sky for one year is over 21 hours. A calibration signal is recorded six times in 24 hours in the form of  "ON-OFF-ON " (calibration step) of a known temperature,  allowing us to equalize the gain in the frequency channels (see details in \citet{Tyulbashev2020}).
	
While searching for new RRAT on LPA LPI, a candidate was recorded in the catalog for further verification, if the signal-to-noise ratio ($S/N$) captured by the program was $\ge7$ after compensation of the dispersion delay \citep{Tyulbashev2018b}. Initially, this $S/N$ level was chosen based on the considerations that these signals are reliably detected visually on dynamic spectra. At the same time, for RRAT which has a defined period ($P_0$) and a dispersion measure
($DM$), it is possible to reprocess the data and purposefully search for weaker pulses. These pulses may be hardly distinguishable or not at all visible on the dynamic spectra, but they will be reliably detected taking the known period into account.
	
%The search for weak pulses from objects with known periods and dispersion measures that we implemented differs from the previously used blind search for new RRAT.
 There are two main problems of transient pulses search. The first one is an interference removal. Most of interferences are of an impulse nature. We compared the found pulses signal-to-noise on $DM=0 pc \cdot cm^{-3}$ and on DM of the transient, which allowed us to remove interferences. The second problem is proper background subtraction.
%The main difference is the way the background signal is subtracted.
In the initial RRAT search, linear approximation was used to subtract the background signal. The time interval at which the subtraction was performed was approximately 15 time seconds. After removing the interference, the noise root dispersion was evaluated over the entire interval. The obtained estimate of the noise root-mean-square deviations was used to determine $S/N \ge 7$ for observed peak signals. At the same time, if a fifteen-second recording still contains interference that lasts more than a few seconds or scintillations from background sources, the resulting noise dispersion estimate will be incorrect, and therefore the $S/N$ estimate will also be incorrect. In particular, the characteristic time scale of scintillations on interplanetary plasma at a frequency of 111 MHz is comparable in duration to the
recording of the RRAT signal (if $DM < 100$~$pc\cdot cm^{-3}$) in the dynamic spectrum. If a single scintillation coincides with the time of arrival of the transient pulse, a situation may arise where the background signal must be subtracted at an interval of few seconds.

The current implementation of the background subtraction algorithm is that the background signal is determined on an array of 32-channel data added with a zero dispersion measure after removing pulse interferences. A power polynomial was fitted into the resulting array. It is believed that this polynomial fits the background signal. Check for polynomials of different powers showed that the most precise fit is for polynomials with a power greater than seven. Illustration of the background subtraction is in the Fig.1.

   \begin{figure}[h]
   \centering
   \includegraphics{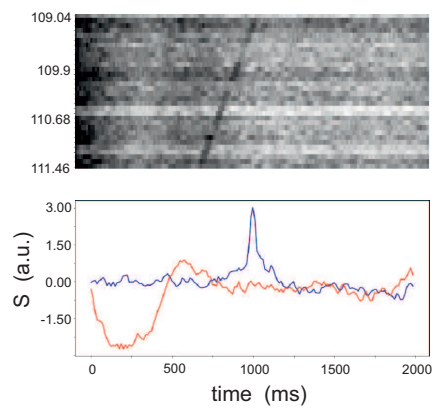}
     \caption{Vertical scale in the upper line of figures on the dynamic spectrum shows the frequencies used in 32-channel observations. The horizontal scale of the lower line of drawings shows the recording time. The upper part of the figure shows the dynamic spectrum, and the lower part of the figure shows the pulse profile with height in arbitrary units (a.u.). The blue color shows the pulse obtained by adding up the channels represented in the dynamic spectra after taking the dispersion measure into considering, and the red color shows the signal added up with the zero dispersion measure. The detected pulse has $S/N=18.3$.
}
       %  \label{Fig1}
   \end{figure}

   \begin{figure*}
%   \centering
   \includegraphics{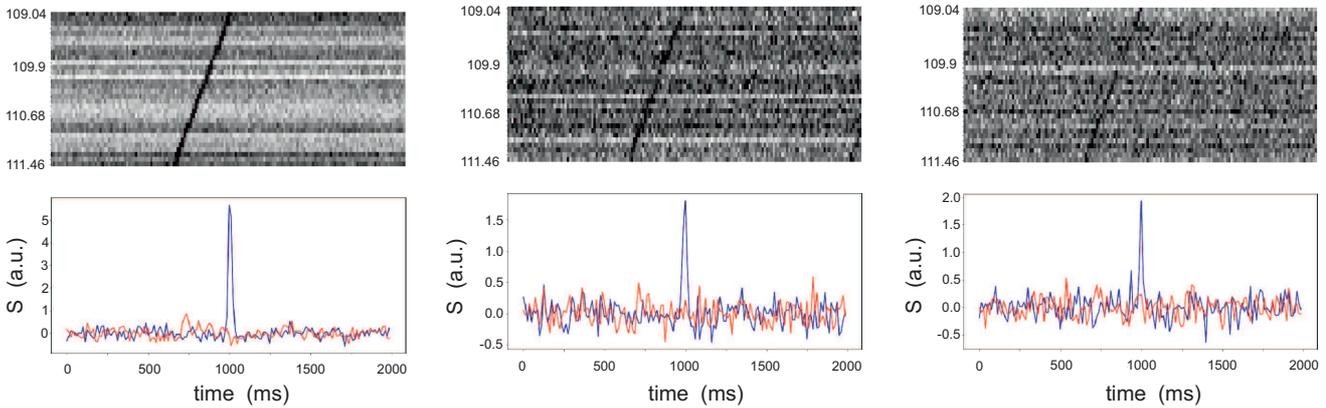}
     \caption{Figure from left to right shows three pulses from RRAT J0139+33, detected in a single observation session and having $S/N=29;10;11$. The horizontal and vertical axes are similar to the axes in Fig.1.}
       %  \label{Fig2}
   \end{figure*}

   At the next step, an array of 32-channel data was added taking the dispersion measure of the studied transient into account and a power polynomial was subtracted from the resulting array. The noise dispersion was determined at an interval of four seconds after excluding the signal from the transient itself.

There were two main criteria for selecting the pulses of the RRAT under study. First, the signal-to-noise ratio of detected pulses $S/N \ge 7$.
Second, the peak height in an array added with a known dispersion measure of RRAT is 1.5 times higher than in an array added with a zero dispersion measure of variance. The second criterion allowed us to get rid of interference that has a duration of more than a second.

      As an illustration of the search algorithms, we present three pulses from J0139+33 detected in the data of August 5, 2017 (see Fig.2).

  \begin{figure}
   \centering
   \includegraphics[width=6cm]{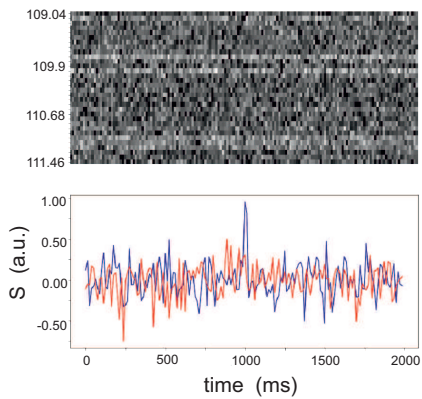}
     \caption{One of the weak pulses from J0139+33. The horizontal and vertical axes are similar to the axes in Fig.1. The detected pulse has $S/N=4.9$.
}
       %  \label{Fig3}
   \end{figure}

   Pulse search was to $S/N=4$. Extremely weak pulses that are still visible on dynamic spectra have $S/N<5$ (see Fig.3). However, with these signal-to-noise ratios, false detections appear, which are practically absent for transients with $S/N > 7$. Therefore, pulses with $S/N< 7$ were excluded during further statistical analysis.
     \section{Results}

      The search for all pulses that fall into the beam of LPA is for RRAT J0139+33, having $P_0=1.2479$~s, $DM=21.23~pc\cdot cm^{-3}$, $D=0.98$~kpc \citep{Sanidas2019} and for the canonical pulsar B0320+39, having $P_0=3.032$~s, $DM=26.18~pc\cdot cm^{-3}$, $D=0.95$~kpc (ATNF catalog \footnote{$https://www.atnf.csiro.au/people/pulsar/psrcat/$}).  Both studied objects are second pulsars, have close dispersion measures, and are located at the same distances.

       Because the LPA LPI is a phased array, the shape of its beam obeys the dependence $(sin(x)/x)^2$. The consequence is that the transient pulse recorded at the top of the emission pattern (the moment the source passes through the meridian) and the same energy pulse recorded on the slope of the emission pattern will have different responses (different heights). The coordinates of the objects under study are known, therefore it is possible to make compensation for the heights of the detected pulses, taking the shape of the LPA beam into account. However, the correction of the beam can be high in the area of the zeros of the beam. Additionally, the pulse coordinates for cases of particularly strong ionospheric scintillations can be shifted by tens of time seconds and, thus, the formal application of the correction to compensate for the height of the detected pulse can increase the pulse height many times and be incorrect. To draw the dependencies, we used pulses found near the maximum of the emission pattern in the time interval $\pm1.5^m$ from the top of the source. The correction of the height of the detected pulses does not exceed $40\%$, and therefore we can expect that the resulting histograms are not distorted.

      A total of 60830 ($S/N \ge 7$) B0320+39 pulsar pulses were detected in 885 observation sessions and 1595 ($S/N \ge 7$) J0139+33 transient pulses in 1669 sessions. On the interval $\pm1.5^m$ from the time of passing through the Meridian of the studied objects, 39377 pulses for PSR B0320+39 were found, and for RRAT J0139+33, 1013 pulses were found. The average frequency of occurrence of pulses from J0139+33 is pulse during 4.9 minutes of recording. The total loss of pulses for B0320+39 is $26\%$, and for J0139+33, it is $99.58\%$. Random check shows that in high-quality records, the number of registered B0320+39 pulsar pulses is almost the same as the number of periods. For illustration, Fig.4 shows a typical records of one of the days with the highest number of registered pulses for B0320+39 and J0139+33.

   \begin{figure*}
   \centering
   \includegraphics[width=\textwidth]{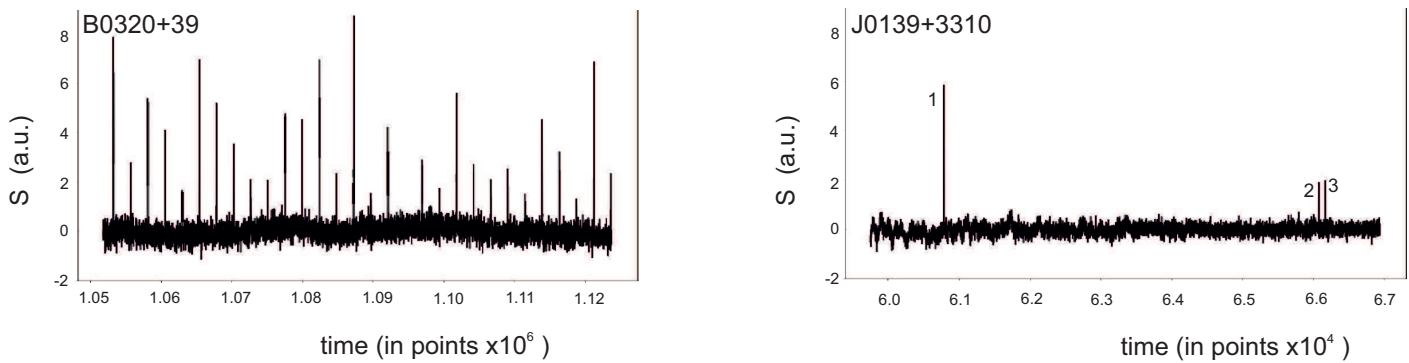}
     \caption{ Figure shows 1.5 minutes of recording after subtracting the background signal and compensating for known dispersion measures for PSR B0320+39 and RRAT J0139+33. Only one pulse is not visible in the record for the pulsar, which is located between the points 109000 and 110000. For the transient, the recording for August 05, 2017 is shown. Three pulses are visible on the recording. The dynamic spectra of these pulses and the average profiles are shown in Fig.1. The time interval between pulses labeled "1" and "3" is 67.44 s, which corresponds to 54 periods of J0139+33. The time interval between pulses labeled "2" and "3" is equal to one period. The vertical axis indicates the intensity in arbitrary units (a.u.) after removing the background signal. On the horizontal axis, the time is counted at points from the beginning of the observation session. One point corresponds to 12.5 ms. On both pictures, the intensity is normalized by the calibration step.
}
       %  \label{Fig3}
   \end{figure*}

    \begin{figure*}
   \centering
   \includegraphics[width=\textwidth]{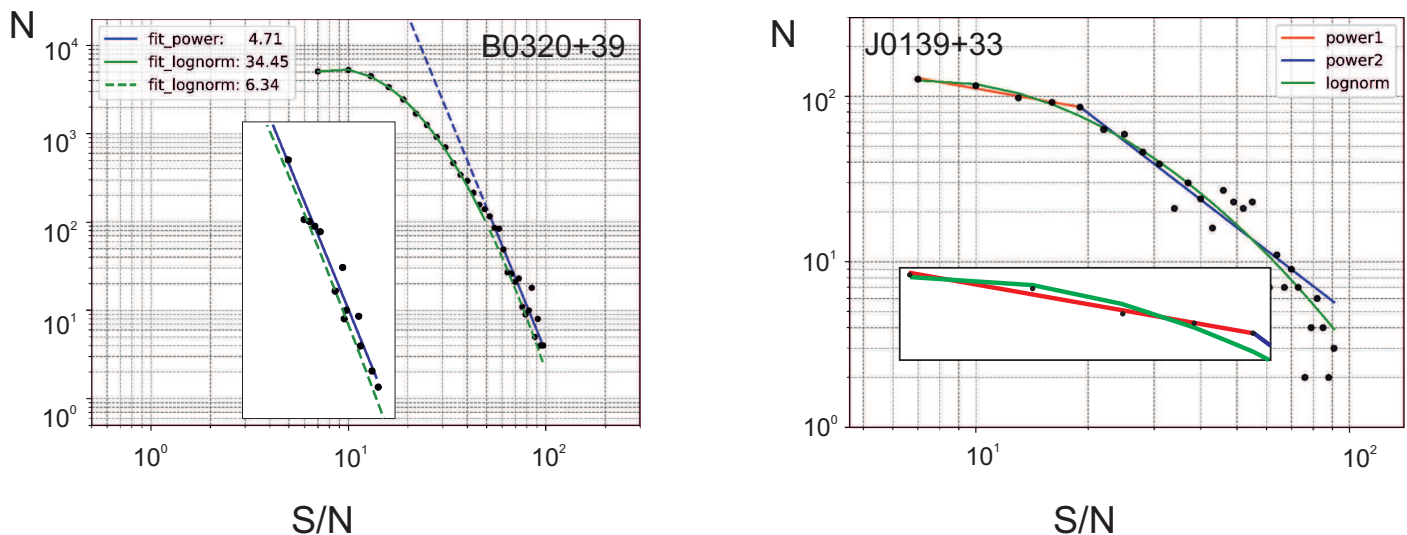}
     \caption{Distribution of peak flux densities in S/N units on the logarithmic scale for the pulsar B0320+39 (left), for transient J0139+33 (right). The vertical axis shows the number of detected pulses, and the horizontal axis shows the signal-to-noise ratio of these pulses. The histograms were drawn in increments of three on the abscissa axis. The figure shows that the distributions are different, despite the similar values of the dispersion measures, distances, and flux densities of the strongest observed pulses in the second pulsar B0320+39 and the second transient J0139+33. The lognormal distribution with probable power tail (see the fragment of the figure) was fitted for B0320+39 and broken power dependence was fitted for J0139+33 (see the fragment of the figure). Standard deviations are shown on the upper corner of B0320+39.
}
       %  \label{Fig4}
   \end{figure*}

      The ratio of peak flux densities of the strongest to the weakest ($S/N= 7$) detected pulses for RRAT J0139+33 is 22 (the strongest pulse S/N=154), and for PSR B0320+39 it is 37.4 (the strongest pulse $S/N =262$).
      
\begin{table}
\caption{Standard deviations and AICc of J0139+33}
\label{table:1}
\begin{tabular}{lll}
\hline
tested model     & std   & AICc      \\
\hline
power(red+blu)   & 4.46    & 92  	 \\
lognorm (green)  & 4.83    & 96   	 \\
\hline
\end{tabular}
\end{table}      

      For all pulses detected in the interval $\pm1.5^m$ of the maximum of the beam, distributions of peak flux densities expressed in signal-to-noise units were drawn (fig.5). It is clear that the distributions for B0320+39 and J0139+33 are different. The most precise fitting with broken power-law dependence with exponents 0.4 and 1.8 was shown for J0139+33. 
      %The lognormal distribution fits well into the data for the pulsar at B0320+39 for low $S/N$, while the power distribution fits well for high $S/N$. The test of most precise fitting with Akaike information criterion (AIC Python programmes \footnote{https://github.com/lmfit/lmfit-py/}) between lognormal and lognormal ({$S/N \le 49$}) plus power distribution ({$S/N \ge 50$}) for B0320+39, and lognormal and two different power distribution for J0139+33 were confirmed our choice.
      We have placed in the Table~1 some results of testing of models using the standard deviation and the Akaike information criterion (AIC\footnote{https://github.com/lmfit/lmfit-py/}) corrected for small sample size (AICc\footnote{https://en.wikipedia.org/wiki/Akaike-information-criterion}) in the cases of lognormal and broken power dependence. The broken power dependence fits better for J0139+33 than lognormal dependence.

      The distribution of pulse energy for PSR B0320+39 was studied earlier at LPA LPI \citep{Kazantsev2018}. In this work \citet{Kazantsev2018} it was shown that giant pulses from this pulsar are unlikely. Into the data for RRAT J0139+33 the broken power distribution fits the most precisely, although a visible spread of points it seems to be associated with a low number of detected pulses.
%    \begin{figure}
%   \centering
%   \includegraphics[width=\textwidth]{t20_fig4.eps}
%     \caption{Distribution of peak flux densities in S/N units on the logarithmic scale for the pulsar B0320+39 (left), for transient J0139+33 (right). The vertical axis shows the number of detected pulses, and the horizontal %axis shows the signal-to-noise ratio of these pulses. The histograms were drawn in increments of 3 on the abscissa axis. The scales on the vertical axis are made the same for easy visual comparison of histograms. The figure shows %that the distributions are different, despite the similar values of the dispersion measures, distances, and flux densities of the strongest observed pulses in the second pulsar B0320+39 and the second transient J0139+33. {\bf The %lognormal distribution with probable power tail (see the fragment of the figure) was fitted for B0320+39 and bimodal power dependence was fitted for J0139+33. Root-mean-square (r.m.s.) deviations are shown on the upper corners of %pictures}.
%}
       %  \label{Fig4}
 %  \end{figure}

     \section{Discussion of results}

       There are a number of signs for giant pulses. It is high peak flux density in pulse in comparision with average flux density of pulsar, power law of pulse distribution, short duration or same times extremely short duration of giant pulse, high degree linear and circular polarization, localization at longitude of average pulse profile and others signs (see, for example \citet{Sutton1971,Kinkhabwala2000,Soglasnov2004,Hankins2007,Kazantsev2018}). Unfortunately with our monitoring observations, we can test only pulse distribution and to estimate peak fluxes of pulses.

      In early works, flux density distributions for RRAT pulses were already developed. The distributions for the four RRAT was obtained in a pioneering work \citet{McLaughlin2006}. For two of the four sources, the number of detected pulses was very low and amounted to 11 and 27 pulses, but the authors believe that the distribution of pulses by energy is described by power function. For RRAT J1317-5759 the distribution is based on 108 pulses, and for RRAT J1819-1458 it is based on 229 pulses. The distribution of pulses by energy was also found to be power function with an exponent of about one.

      However, other authors \citep{Keane2010,Cui2017,Mickaliger2018,Meyers2019} when drawing pulse energy distributions for almost 30 RRATs, have found that in general the distributions are close to the lognormal or the sum of two lognormal distributions. Moreover, both transients defined in the paper \citet{McLaughlin2006} as transients with a power distribution of pulses over energies are marked in the paper \citet{Mickaliger2018} as transients whose distribution consists of two lognormal distributions. The authors associate the lognormal distribution, which has a maximum at high flux densities, with possible flash activity in these RRATs. In the paper \citet{Mickaliger2018}, it is noted that there are known flashing pulsars with a similar (double-peak) distribution. In \citet{Cui2017}, it is noted that for some of the transients, the distribution is better described by the sum of the lognormal and power distributions.

      The study of the energy distribution of pulses for canonical pulsars was carried out in the works \citet{Burke-Spolaor2012,Mickaliger2018}. In these papers, distributions were drawn for more than 300 pulsars and it was shown that for most of them the dependence is well described by the lognormal distribution or the sum of the lognormal and power distributions. That is, in general, the distributions for RRATs and for pulsars are similar. At the same time, the distribution is lognormal on the side of low flux densities, and a power dependence is often added to the tail of the distribution.

      %However, we have not found any publications with the study of the energy distribution of pulses for RRAT J0139+33.
      The dependence mentioned in the previous section for J0139+33 is described by a power-law. A power dependence in the pulse distribution is observed for a millisecond pulsar in the Crab nebula. The pulsar in the Crab nebula is the most famous pulsar with giant pulses. The paper \citet{Mickaliger2012} presents 13 estimates of the exponent obtained by different authors for the distribution of pulses by energy at frequencies from 112 MHz to 4850 MHz. The estimation of the exponent varies from 2.1 to 4.2, and its median value falls to 2.8. For seconds pulsars, a similar behavior is demonstrated by the B0950+08 pulsar in which giant pulses are observed \citep{Singal2001,Kazantsev2018}. The broken power-law dependence is also shown for three pulsars (B0531+21, B0950+08, B1237+25) with giant pulses \citep{Smirnova2012,Kazantsev2017}.
      %For {\bf B}0950+08 the energy distribution of pulses is described by power distribution over a wide range of flux densities. The exponent is estimated as 1.9 \citep{Kazantsev2018}.
      %Thus, the energy distribution of pulses for the RRAT J0139+33 studied by us is typical for giant pulses.

%      {\bf The distribution of pulses for B0320+39 at frequency 111 MHz was builded in paper \citet{Kazantsev2018}. Authors were found lognormal law for pulses but give not the characteristics of this distribution. Publications with the study of the energy distribution of pulses for RRAT J0139+33 did not found. The least square method was used for testing of lognormal (eq.1) and power (eq.2) dependieses.

 %   \begin{equation}
  %  f(x) = \frac{a}{x} \times e^{-\frac{(ln(x)-x_o)^2}{2\sigma^2}},
   % \label{eq1}
    %\end{equation}

    %where

%    \begin{equation}
 %   f(x)=a_1\times x^{-b},
  %  \label{eq2}
   % \end{equation}

    %where

%    We found for PSR B0320+39 lognormal distribution of pulses before $S/N=30-40$ and power distribution of pulses with $S/N \ge 50$. The distribution pulses of RRAT J0139+33 can be described by lognormal or bimodal power law (see table.??), but r.m.s. for bimodal power low is less than for lognormal distribution. In papers \citet{Smirnova2012}, \citet{Kazantsev2017} bimodal distribution of giant pulses observed for 3 pulsars. Thus, the energy distribution of pulses for the RRAT J0139+33 studied by us is typical for giant pulses.}

      The hypotheses mentioned in the introduction can be divided into two types. RRATs are pulsars that do not have periodical emission between the observed sporadic pulses, or RRATs are pulsars that have very weak periodical emission between the sporadic pulses.

     The search for regular emission at a frequency of 111 MHz of 16 rotating transients discovered at 1400 MHz was made on the LPA LPI radio telescope in 2010-2012 \citep{Losovsky2014}. Regular emission was not detected in individual sessions. However, a search using the Fast Folding Algorithm (FFA) followed by averaging all observation sessions revealed weak regular emission of most of the studied RRATs. The estimated peak flux density of detected RRATs ranges from 20 mJy to 90 mJy. The average peak flux density was 45 mJy.

      %Here are additional considerations that suggest that all the pulses of J0139+33 we observe are giant. We will involve known data that allow us to consider the question of regular (pulsar) emission of a rotating transient.
        Can we estimate the flux density of J0139+33? Indirect estimates of the absence of periodic emission of J0139+33 can be derived from the work on the search for pulsars in the summed up power spectra at a frequency of 111 MHz \citep{Tyulbashev2017,Tyulbashev2020}. This work used six-channel data with a total observation bandwidth of 2.5 MHz and a sampling time of 0.1 s. The total increase in signal to noise ratio over the four-year interval was approximately 30 times. No harmonics corresponding to the transient period were found in the summed power spectra. This means that there was no regular periodic emission from the direction of J0139+33.  Because we know the exact coordinate of the transient and the pulse duration, we can give a rough estimate of the integral flux density, which is based on the assumption that the root-mean-square deviations of noise during observations are 0.1 Jy and that the transient is a canonical pulsar with constant, albeit weak, emission. According to table 1 from the work \citep{Tyulbashev2016}, showing the expected sensitivity in a single session of observations on the LPA LPI as 15-20 mJy for the worst case at the optimal reading time and $S/N>7$. The upper limit of the integral flux density from J0139+33 after summation of power spectra can be given as $S_\mathrm{int}< 2$~mJy at the observation frequency of 111 MHz.

      RRAT J0139+33 was also detected by studying its individual pulses in a survey on the search for pulsars on the LOFAR ({\bf LO}w {\bf F}requency {\bf A}rray) system \citep{Sanidas2019}. No periodic emission was detected for the transient when using standard search programs. According to fig.3 in this paper, the sensitivity of LOFAR survey observations for pulsars located in the extragalactic plane and having a dispersion measure of less than $100~pc \cdot cm^{-3}$ is estimated as 1.2 mJy at the observation frequency of 135 MHz.

      Thus, both observations on the LPA LPI and observations on LOFAR indicate that the integral flux density of J0139+33 does not exceed 2 mJy at the frequency of 111 MHz. Let's suppose that J0139+33 is a canonical pulsar, its flux density $S_\mathrm{int}=2$~mJy, and the half-width of the average profile coincides with the half-width of the observed transient pulses. According to \citep{Sanidas2019} the pulsar period $P=1.2479$~ms, and the half-width of the profile $w_{50}=25$~ms. Then the upper estimate of the peak flux density in the average profile $S_\mathrm{peak}=(1.2479/0.025) \times 2mJy\le 100$~mJy. Let's convert the signal-to-noise ratio for the strongest detected pulses into units of flux density using data on the fluctuation sensitivity of the LPA LPI antenna \citep{Tyulbashev2016}. The peak flux density will be equal to 46 Jy ($S/N=154$). Thus, the observed intensity of the strongest observed pulse is at least 460 times higher than the expected average pulse of J0139+33, if the rotating transient still has a weak regular emission. This excess of energy in strong pulses compared to the average pulsar pulses is also characteristic of giant pulses. If weak regular emission will be detected in the future (see work of \citet{Losovsky2014}), the conclusion about the detection of giant pulses J0139+33 will be strengthened.

     % {\bf Generally speaking there is no good definition of a giant pulse. }

      %Let's consider the hypotheses mentioned in the Introduction. The hypothesis that the observed pulses of rotating transients are giant pulses of ordinary pulsars that are invisible due to their low integral flux density \citep{Weltevrede2006} fits well for RRAT J0139+33.
      Regular emission from J0139+33 has not yet been detected, and therefore it is impossible to be completely sure that J0139+33 is weak pulsar with giant pulses.
      %this hypothesis uniquely explains the studied RRAT. We can only say that the luminosity of J0139+33 will be many times less than the luminosity of B0320+39.
      Detecting regular emission from J0139+33 will be a good proof that all the observed transient pulses are gigantic.

     % {\bf ???} The hypothesis that the absence of periodic emission is associated with long (extreme) nulling periods \citep{Zhang2007} implies that the energy distribution in the observed pulses is lognormal. This hypothesis is not confirmed for the studied transient.

     \section{Conclusion}

      Based on observations at a frequency of 111 MHz, the distribution of peak flux densities for J0139+33 in units of $S/N$ was described, and it turned out to be broken power-law dependence. Regular emission of J0139+33 is not detected up to the integral flux densities of 1-2 mJy. It is shown that probably J0139+33 is a rotating transient with giant pulses.

\begin{acknowledgements}
The authors express their gratitude for the help in the course of this work to A. N. Kazantsev, V. A. Potapov, M. A. Kitaeva, and L. B. Potapova. We thank the anonymous referee for a number of helpful suggestions that improved the paper.
\end{acknowledgements}

\bibliographystyle{aa} % style aa.bst
\bibliography{serg1} % your references Yourfile.bib

\begin{thebibliography}{34}
\expandafter\ifx\csname natexlab\endcsname\relax\def\natexlab#1{#1}\fi

\bibitem[{{Burke-Spolaor} \& {Bailes}(2010)}]{Burke-Spolaor2010}
{Burke-Spolaor}, S. \& {Bailes}, M. 2010, \mnras, 402, 855

\bibitem[{{Burke-Spolaor} {et~al.}(2012){Burke-Spolaor}, {Johnston}, {Bailes},
  {Bates}, {Bhat}, {Burgay}, {Champion}, {D'Amico}, {Keith}, {Kramer}, {Levin},
  {Milia}, {Possenti}, {Stappers}, \& {van Straten}}]{Burke-Spolaor2012}
{Burke-Spolaor}, S., {Johnston}, S., {Bailes}, M., {et~al.} 2012, \mnras, 423,
  1351

\bibitem[{{Cui} {et~al.}(2017){Cui}, {Boyles}, {McLaughlin}, \&
  {Palliyaguru}}]{Cui2017}
{Cui}, B.~Y., {Boyles}, J., {McLaughlin}, M.~A., \& {Palliyaguru}, N. 2017,
  \apj, 840, 5

\bibitem[{{Hankins} \& {Eilek}(2007)}]{Hankins2007}
{Hankins}, T.~H. \& {Eilek}, J.~A. 2007, \apj, 670, 693

\bibitem[{Hewish {et~al.}(1968)Hewish, Bell, Pilkington, Scott, \&
  Collins}]{Hewish1968}
Hewish, A., Bell, S., Pilkington, J., Scott, P., \& Collins, R. 1968, \nat,
  217, 709

\bibitem[{{Kazantsev} \& {Potapov}(2017)}]{Kazantsev2017}
{Kazantsev}, A.~N. \& {Potapov}, V.~A. 2017, Astronomy Reports, 61, 747

\bibitem[{{Kazantsev} \& {Potapov}(2018)}]{Kazantsev2018}
{Kazantsev}, A.~N. \& {Potapov}, V.~A. 2018, Research in Astronomy and
  Astrophysics, 18, 097

\bibitem[{Keane {et~al.}(2010)Keane, Ludovici, Eatough, Kramer, Lyne,
  McLaughlin, \& Stappers}]{Keane2010}
Keane, E., Ludovici, D., Eatough, R., {et~al.} 2010, \mnras, 401, 1057

\bibitem[{Keane \& Kramer(2008)}]{Keane2008}
Keane, E.~F. \& Kramer, M. 2008, \mnras, 391, 2009

\bibitem[{Keane {et~al.}(2011)Keane, Kramer, Lyne, Stappers, \&
  McLaughlin}]{Keane2011}
Keane, E.~F., Kramer, M., Lyne, A.~G., Stappers, B.~W., \& McLaughlin, M.~A.
  2011, \mnras, 415, 3065

\bibitem[{{Kinkhabwala} \& {Thorsett}(2000)}]{Kinkhabwala2000}
{Kinkhabwala}, A. \& {Thorsett}, S.~E. 2000, \apj, 535, 365

\bibitem[{{Li}(2006)}]{Li2006}
{Li}, X.-D. 2006, \apjl, 646, L139

\bibitem[{{Lorimer} \& {Kramer}(2004)}]{Lorimer2004}
{Lorimer}, D.~R. \& {Kramer}, M. 2004, {Handbook of Pulsar Astronomy}, Vol.~4

\bibitem[{{Losovsky} \& {Dumsky}(2014)}]{Losovsky2014}
{Losovsky}, B.~Y. \& {Dumsky}, D.~V. 2014, Astronomy Reports, 58, 537

\bibitem[{{Luo} \& {Melrose}(2007)}]{Luo2007}
{Luo}, Q. \& {Melrose}, D. 2007, \mnras, 378, 1481

\bibitem[{McLaughlin {et~al.}(2006)McLaughlin, Lyne, Lorimer, Kramer, Faulkner,
  Manchester, \& Cordes}]{McLaughlin2006}
McLaughlin, M., Lyne, A., Lorimer, D., {et~al.} 2006, \nat, 439, 817

\bibitem[{{Meyers} {et~al.}(2019){Meyers}, {Tremblay}, {Bhat}, {Shannon},
  {Ord}, {Sobey}, {Johnston-Hollitt}, {Walker}, \& {Wayth}}]{Meyers2019}
{Meyers}, B.~W., {Tremblay}, S.~E., {Bhat}, N.~D.~R., {et~al.} 2019, \pasa, 36,
  e034

\bibitem[{{Michilli} {et~al.}(2020){Michilli}, {Bassa}, {Cooper}, {Hessels},
  {Kondratiev}, {Sanidas}, {Stappers}, {Tan}, {van Leeuwen}, {Cognard},
  {Grie{\ss}meier}, {Lyne}, {Verbiest}, \& {Weltevrede}}]{Michilli2020}
{Michilli}, D., {Bassa}, C., {Cooper}, S., {et~al.} 2020, \mnras, 491, 725

\bibitem[{{Mickaliger} {et~al.}(2018){Mickaliger}, {McEwen}, {McLaughlin}, \&
  {Lorimer}}]{Mickaliger2018}
{Mickaliger}, M.~B., {McEwen}, A.~E., {McLaughlin}, M.~A., \& {Lorimer}, D.~R.
  2018, \mnras, 479, 5413

\bibitem[{{Mickaliger} {et~al.}(2012){Mickaliger}, {McLaughlin}, {Lorimer},
  {Langston}, {Bilous}, {Kondratiev}, {Lyutikov}, {Ransom}, \&
  {Palliyaguru}}]{Mickaliger2012}
{Mickaliger}, M.~B., {McLaughlin}, M.~A., {Lorimer}, D.~R., {et~al.} 2012,
  \apj, 760, 64

\bibitem[{{Sanidas} {et~al.}(2019){Sanidas}, {Cooper}, {Bassa}, {Hessels},
  {Kondratiev}, {Michilli}, {Stappers}, {Tan}, {van Leeuwen}, {Cerrigone},
  {Fallows}, {Iacobelli}, {Orr{\'u}}, {Pizzo}, {Shulevski}, {Toribio}, {ter
  Veen}, {Zucca}, {Bondonneau}, {Grie{\ss}meier}, {Karastergiou}, {Kramer}, \&
  {Sobey}}]{Sanidas2019}
{Sanidas}, S., {Cooper}, S., {Bassa}, C.~G., {et~al.} 2019, \aap, 626, A104

\bibitem[{Shishov {et~al.}(2016)Shishov, Chashei, Oreshko, Logvinenko,
  Tyul'bashev, Subaev, Svidskii, Lapshin, \& Dagkesamanskii}]{Shishov2016}
Shishov, V., Chashei, I., Oreshko, V., {et~al.} 2016, Astronomy Reports, 60,
  1067

\bibitem[{{Singal}(2001)}]{Singal2001}
{Singal}, A.~K. 2001, \apss, 278, 61

\bibitem[{{Smirnova}(2012)}]{Smirnova2012}
{Smirnova}, T.~V. 2012, Astronomy Reports, 56, 430

\bibitem[{{Soglasnov} {et~al.}(2004){Soglasnov}, {Popov}, {Bartel}, {Cannon},
  {Novikov}, {Kondratiev}, \& {Altunin}}]{Soglasnov2004}
{Soglasnov}, V.~A., {Popov}, M.~V., {Bartel}, N., {et~al.} 2004, \apj, 616, 439

\bibitem[{{Sutton} {et~al.}(1971){Sutton}, {Staelin}, \& {Price}}]{Sutton1971}
{Sutton}, J.~M., {Staelin}, D.~H., \& {Price}, R.~M. 1971, 46, 97

\bibitem[{Tyul'bashev {et~al.}(2017)Tyul'bashev, Tyul'bashev, Kitaeva,
  Chernyshova, Malofeev, Chashei, Shishov, Dagkesamanskii, Klimenko, Nikitin,
  \& Nikitina}]{Tyulbashev2017}
Tyul'bashev, S., Tyul'bashev, V., Kitaeva, M., {et~al.} 2017, Astronomy
  Reports, 61, 848

\bibitem[{Tyul'bashev {et~al.}(2016)Tyul'bashev, Tyul'bashev, Oreshko, \&
  Logvinenko}]{Tyulbashev2016}
Tyul'bashev, S., Tyul'bashev, V., Oreshko, V., \& Logvinenko, S. 2016,
  Astronomy Reports, 60, 220

\bibitem[{{Tyul'bashev} {et~al.}(2020){Tyul'bashev}, {Kitaeva}, {Tyul'basheva},
  {Malofeev}, \& {Tyul'bashev}}]{Tyulbashev2020}
{Tyul'bashev}, S.~A., {Kitaeva}, M.~A., {Tyul'basheva}, G.~E., {Malofeev},
  V.~M., \& {Tyul'bashev}, V.~S. 2020, Astronomy Reports, 64, 526

\bibitem[{{Tyul'bashev} {et~al.}(2018{\natexlab{a}}){Tyul'bashev},
  {Tyul'bashev}, \& {Malofeev}}]{Tyulbashev2018b}
{Tyul'bashev}, S.~A., {Tyul'bashev}, V.~S., \& {Malofeev}, V.~M.
  2018{\natexlab{a}}, \aap, 618, A70

\bibitem[{{Tyul'bashev} {et~al.}(2018{\natexlab{b}}){Tyul'bashev},
  {Tyul'bashev}, {Malofeev}, {Logvinenko}, {Oreshko}, {Dagkesamanskii},
  {Chashei}, {Shishov}, \& {Bursov}}]{Tyulbashev2018a}
{Tyul'bashev}, S.~A., {Tyul'bashev}, V.~S., {Malofeev}, V.~M., {et~al.}
  2018{\natexlab{b}}, Astronomy Reports, 62, 63

\bibitem[{{Wang} {et~al.}(2007){Wang}, {Manchester}, \& {Johnston}}]{Wang2007}
{Wang}, N., {Manchester}, R.~N., \& {Johnston}, S. 2007, \mnras, 377, 1383

\bibitem[{{Weltevrede} {et~al.}(2006){Weltevrede}, {Stappers}, {Rankin}, \&
  {Wright}}]{Weltevrede2006}
{Weltevrede}, P., {Stappers}, B.~W., {Rankin}, J.~M., \& {Wright}, G.~A.~E.
  2006, \apjl, 645, L149

\bibitem[{{Zhang} {et~al.}(2007){Zhang}, {Gil}, \& {Dyks}}]{Zhang2007}
{Zhang}, B., {Gil}, J., \& {Dyks}, J. 2007, \mnras, 374, 1103

\end{thebibliography}

\end{document}